\documentclass[showpacs,reprint,prb,amsmath,amssymb,floatfix,superscriptaddress]{revtex4-1}
\usepackage{graphicx}
\usepackage{amsmath}
\usepackage{amssymb}
\usepackage{amsfonts}
\usepackage{bm}
\usepackage{color}

\begin{document}

\title{Work Extraction and Landauer's Principle in a Quantum Spin Hall
Device}
\author{A. Mert Bozkurt}
\author{Bar{\i}\c{s} Pekerten}
\author{ \.{I}nan\c{c} Adagideli}
\email{adagideli@sabanciuniv.edu}
\affiliation{Faculty of Engineering and Natural Sciences, Sabanci
University, Orhanli-Tuzla, Istanbul, Turkey}

\date{\today }

\begin{abstract}
Landauer's principle states that erasure of each bit of information in
a system requires at least a unit of energy $k_B T \ln 2$ to be
dissipated. In return, the blank bit may possibly be utilized to
extract usable work of the amount $k_B T \ln 2$, in keeping with the
second law of thermodynamics.  While in principle any collection of
spins can be utilized as information storage, work extraction by
utilizing this resource in principle requires specialized engines that
are capable of using this resource.  In this work, we focus on heat and
charge transport in a quantum spin Hall device in the presence of a
spin bath. We show how a properly initialized nuclear spin subsystem can be used as
a memory resource for a Maxwell's Demon to harvest available heat
energy from the reservoirs to induce charge current that can power an
external electrical load.  We also show how to initialize the nuclear spin subsystem
using applied bias currents which necessarily dissipate energy, hence
demonstrating Landauer's principle. This provides an alternative method
of ``energy storage'' in an all-electrical device. We finally propose a
realistic setup to experimentally observe a Landauer erasure/work
extraction cycle.

\end{abstract}

\pacs{74.78.Na, 85.75.-d, 72.10.-d}

\maketitle

\section{Introduction}


According to Landauer's principle, erasure of one bit of information
requires an amount of heat greater than $k_B T \ln 2$ to be dissipated
~\cite{REF:Landauer1961, REF:Bennett1982}. The principle ensures that
the second law of thermodynamics is obeyed as a blank bit is utilized
to extract work by an amount $k_B T \ln 2$ from the environment. The
``engine'' that is capable of this extraction is sometimes called a
``Maxwell's Demon'' (MD), referring to the thought experiment proposed
by Maxwell in 1871~\cite{REF:MaxwellDemon}. While interest in MD and
Landauer's principle from the point of view of fundamental physics
never faded~\cite{REF:Lloyd1997, REF:Scully2001, REF:Kieu2004,
REF:Vaccaro2011, REF:Barnett2013, REF:Croucher2017}, promise of highly
efficient engines that operate in the nano-domain~\cite{REF:DattaBook,
REF:vonOppen2013, REF:Arrachea2015} as well as alternative methods of
energy storage gave a recent impetus to research on the physics of MD
both experimentally (using colloidal particles~\cite{REF:Toyabe2010,
REF:Berut2012, REF:Roldan2014}, photonic
systems~\cite{REF:Vidrighin2016, REF:Ciampini2017}, NMR
systems~\cite{REF:Camati2016, REF:Peterson2016}, single electron
transistors~\cite{REF:Koski2014v1, REF:Koski2014v2, REF:Koski2015,
REF:Chida2017}, cavity QED with superconducting
qubits~\cite{REF:Cottet2017}) and theoretically~\cite{REF:Anders2011,
REF:Mandal2013, REF:Barato2013, REF:Strasberg2013, REF:Horowitz2013,
REF:Deffner2013, REF:Park2013, REF:Rossnagel2014, REF:Strasberg2014,
REF:Gemmer2015, REF:Pekola2016, REF:Kutvonen2016, REF:Lebedev2016,
REF:Campisi2017, REF:Elouard2017, REF:Platero2017}. Despite the
multitudinous platforms in which MD action is theorized or
demonstrated, scalability remains an issue.

In this manuscript, we propose and investigate a new MD implementation
that harvests thermal energy from the electronic environment and
converts it to electrical work using a quantum spin Hall insulator
(QSHI). As a memory resource we use the available nuclear spins present in the device and/or
magnetic impurities introduced via doping. QSHIs feature an insulating
bulk and a pair of counter-propagating gapless  spin-momentum locked
helical edge states that are topologically protected from
backscattering under time-reversal symmetry (TRS)~\cite{REF:Konig2008}
(see Fig.~\ref{FIG:QSH_Nuclear}). First predicted to exist in graphene
nanoribbons~\cite{REF:Kane2005a, REF:Kane2005b}, they were later
experimentally found in HgTe/CdTe quantum wells
(QWs)~\cite{REF:Konig2007, REF:Bernevig2006} as well as in InAs/GaSb QW
structures~\cite{REF:Liu2008, REF:Du2015}. The TRS prohibiting the
backscattering of the edge states is broken by the presence of nuclear
or impurity spins. This backscattering shows up as extra dissipation,
lowering the expected quantized conductance of the QSHI
edge~\cite{REF:Lunde2012, REF:Cheianov2013, REF:DelMaestro2013,
REF:Kimme2016, REF:Loss2017}. Here we show another salient feature of
such scattering: an initial state of polarized nuclear spins (blank
memory) drives an electric current. Thus nuclear/impurity spins act as
a memory resource of a MD that converts heat from the environment into
electrical work.

We show below that for the heat harvesting operation of our engine, no
energy exchange between nuclear and electronic systems is necessary; in
fact, the nuclear spins are degenerate in our system, forming a
non-energetic (pure) memory. Hence this is an alternative way for
energy storage (Fig.~\ref{FIG:QIE_circuit}) that is protected from
undesired explosive discharges. The total energy needed to reset the
``memory'' (or, in other words, recharge the device) by fully
polarizing nuclear spins exceeds the extracted energy, in agreement
with the second law of thermodynamics and Landauer's principle. We also
provide a method to generate such a nuclear spin polarization,
completing the discharge-recharge cycle of the quantum information
engine (QIE) (Fig.~\ref{FIG:QIE_circuit}). We note that each nucleus
with nonzero nuclear spin coupling to the electron spin in the QSHI
edge contributes to the MD memory, hence the MD memory size here could
be several orders of magnitude larger compared to those that were
reported in the literature, thus solving the scaling problem for heat
harvesting engines. Furthermore, our estimates show that equivalent
energy/power density of our proposed engine compares favourably with
conventional energy storage such as supercapacitors.

\section{Nuclear spins in Quantum Spin Hall Insulators}

\begin{figure}[tb]
\includegraphics[width=.95\columnwidth]{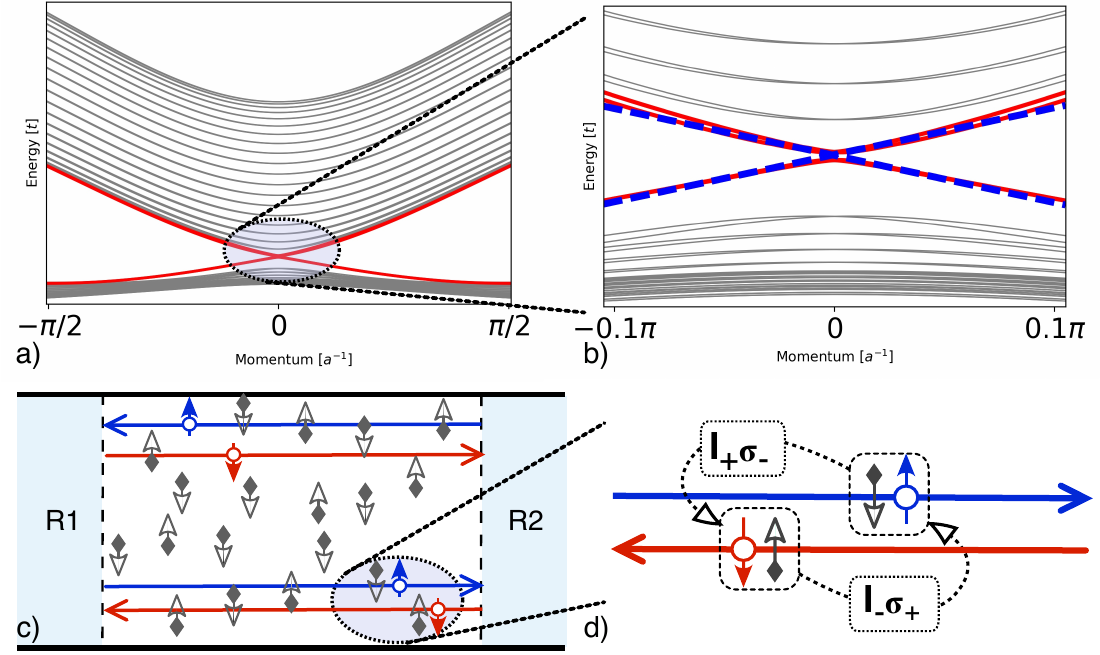}
\caption{(Color online) QSHI with nuclear spins and electron-nuclear
spin flip interaction. (a) The band structure of a typical QSHI system
(using the BHZ model with tight-binding approximation). Red lines
represent the edge states. (b) The band structure of the simplified
Hamiltonian $h_\textrm{eff}$ projected to a single edge (dashed blue
lines). (c) Schematic description of the QSHI system with the edge
currents interacting with the nuclear spins in the system, with the
diamonds representing nuclear spins. (d) The spin flip interaction with
the nuclear spins that form the MD.}\label{FIG:QSH_Nuclear}
\end{figure}

We now describe the basic model of our MD
implementation. The effective dynamics of electrons and holes in QSHI
materials is well described by the Bernevig-Hughes-Zhang (BHZ)
Hamiltonian~\cite{REF:Bernevig2006}:
\begin{align}
\label{EQN:BHZ_general}
\mathcal{H}_{\textrm{BHZ}} = \epsilon_k \, \sigma_0\tau_z + E_k\, \sigma_0\tau_0 +
A\,(k_x \,\sigma_z \tau_x - k_y  \, \sigma_0\tau_y),
\end{align}
where $\epsilon_k = M-Bk^2$, $E_k=C-D k^2$ and $M$, $A$, $B$, $C$, $D$ are the
material parameters. The BHZ Hamiltonian acts on the envelope
wave functions $(\psi_{+,E},\psi_{-,E},\psi_{+,H},\psi_{-,H})^T$, where $\sigma=\pm$ denote the
spin, $\tau=E,H$ denote the electron-hole degrees of freedom and
$\sigma_\alpha$ and $\tau_\alpha$ ($i\in\{x,y,z \})$ are the Pauli matrices
that act in spin and electron-hole spaces
respectively. (We also define the corresponding unit matrices $\sigma_0$ and $\tau_0$.)
In order to describe the coupling to nuclear spins we also need the
full wavefunction which includes the lattice--periodic factors~\cite{REF:Lunde2013}:
\begin{equation}
\Psi(\vec{r}) = \sum_{\sigma,\tau} \psi_{\sigma,\tau}(\vec{r}) u_{\sigma,\tau}(\vec{r}).
\end{equation}
In this description, the various two-dimensional QSHI QW
structures differ only in their material and effective
parameters~\cite{REF:Liu2013}, while the main edge state physics
remains the same (Fig.~\ref{FIG:QSH_Nuclear}a).
The low energy excitations in
the topological phase are localized to the
edges of the system. These excitations are called helical edge states and their
wavefunctions have the general form :
\begin{equation}\label{EQN:EdgeStateElectronicWavefunctions}
\psi_{\sigma,\tau}^\rightleftharpoons(\vec{r}) = \xi(\vec{r}_\perp) \phi^\rightleftharpoons(x),
\end{equation}
where the superscript $\rightleftharpoons$ denotes the edge states of
different chiralities. In the absence of spin-orbit coupling, the
chiralities are specified by their spin so that
$\psi_{\sigma,\tau}^\rightleftharpoons \propto
\delta_{\rightleftharpoons,\sigma}$. (Note that in the presence of
spin-orbit coupling, the spin axes becomes position
dependent~\cite{REF:Adagideli2012}.) We next project the electronic
Hilbert space to the space spanned by the edge states given in
Eq.~\ref{EQN:EdgeStateElectronicWavefunctions} above, obtaining the
projected electron Hamiltonian which reads
$h_{\text{bot(top)}}^\textrm{eff} = \mp i\,\hbar
v_F\partial_x\sigma_z$. Here, $v_F$ is the Fermi velocity of the
effective edge state and $\pm$ signs refer to the top (+) and the
bottom (-) edge (see Fig.~\ref{FIG:QSH_Nuclear}b).

The second important element in our QIE is the nuclear spin subsystem
that forms the ``memory'' of the MD that operates on electron-hole
dynamics via their spins. We model the interaction between the spins of
the nuclei and the spins of the electrons by the Fermi contact
hyperfine interaction~\cite{REF:Slichter1990}, which is given by
\begin{align}\label{EQN:SlichterHyperfine}
H_\textrm{hf} &= v_0 \sum_{i=1}^{N} A_{i}\delta(\vec{r}-\vec{R}_i)
\vec{I_{i}}\cdot \vec{\sigma},
\end{align}
where $\vec{\sigma}$ are the electron spin operators, $v_0$ is the
volume of the unit cell of the corresponding QSHI component material,
$\vec{I_i}$ is the nuclear spin operator at position $\vec{R}_i$ and
$A_i$ is the hyperfine coupling energy.
The total low energy
effective Hamiltonian including the hyperfine interaction projected to the edge states
is then given by
\begin{align}\label{EQN:ModelHamiltonians}
H_{\text{bot(top)}} &= \big(\mp i\hbar v_F\partial_x+\lambda M_z(x)\big) \sigma_z\nonumber\\
H_\textrm{s-flip} &= \sum_{i=1}^{N} \frac{\lambda_i}{2}  \delta(x-x_i)
\big(I_{i+}\sigma_- + I_{i-}\sigma_+\big),
\end{align}
where $x$ denotes the position along the edge in consideration.
In this effective Hamiltonian given in Eq.~ \eqref{EQN:ModelHamiltonians}, the electrons interact with all the nuclear spins within the cross section of the helical edge states, which we denote as S.
We further assume
for simplicity that the effective hyperfine coupling $\lambda_i = \lambda$ is constant for
all sites, which does not qualitatively alter the physics, and estimate its value as
$\lambda = A_0 v_0/S$, where $A_0$ is the average value of $A_i$. In anticipation of dynamically polarized nuclear states that we will consider below,  we have also introduced $M_z(x)$ as
the $z$-component of the Overhauser field~\cite{REF:Slichter1990}.
We
note that $M_z(x)$ can be gauged away via
$H_{\text{bot(top)}}\rightarrow U H_{\text{bot(top)}}\,U^\dagger$ with
$U=\exp \big(\frac{\pm i \lambda}{\hbar v_{F}} \int ^x M_{z}(x') dx'
\big)$.

We note that because the
spin of the electron and its momentum is completely locked, as
$H_\textrm{s-flip}$ flips the spin of the edge electron and one nuclear
spin, it also causes backscattering (see Fig.~\ref{FIG:QSH_Nuclear}d).
We assume that the temperature of the system is higher than
$T^*$, below which RKKY and other nuclear correlation
effects become important. $T^*$ is estimated to
be around $100$ mK or less~\cite{REF:Loss2017}.
The Fermi contact interaction process runs in competition with other
processes that affect the nuclear spins, mainly the quadrupole
interaction causing spin-flip between nuclear spins. However, the rate
of this interaction is orders of magnitude smaller than the spin-flip rate from the coupling
between nuclear and electronic spins~\cite{REF:Dzhioev2007}.
\section{Charging/discharging cycle of QIE}


In this section, we describe the charging/discharging
(or alternatively erasure/work extraction) operation. In the charging
phase, we apply a charge current, which without loss of generality we
assume to be flowing to the left, leading to more right movers than
left movers, and hence to more right to left backscattering. In the
bottom edge, right movers are spin up electrons. Thus the
backscattering creates up-nuclear spins from spin-flip scattering. In
the top edge right movers are spin down electrons hence the
backscattering creates down-nuclear spins (see
Fig.~\ref{FIG:QIE_circuit}b). This process polarizes the nuclear spins
until a certain net bias-dependent value is
reached~\cite{REF:Lunde2012, REF:DelMaestro2013}. This is the process
of dynamical nuclear polarization for the quantum spin Hall edges,
well-known in other contexts such as spin injection from
ferromagnets~\cite{REF:Datta1990, REF:Schmidt2000}. We stress that
under a current bias, opposite edges are driven towards opposite
polarization values.

More importantly, the reverse process is also possible: fully polarized
nuclear spins near a QSHI edge drive a charge current (see
Fig.~\ref{FIG:QIE_circuit}c). This is the discharging phase. Consider a
nonzero initial nuclear spin polarization with opposite signs in
opposite edges (caused by, say, the driving current described above)
and for simplicity assume zero applied voltage bias. Now there are more
up[down]-nuclear spins than down[up]-nuclear spins in the bottom[top]
edge, hence there are more down[up]-spins flipped to up[down]-spins in
the bottom[top] edge, leading to an imbalance of left movers relative
to right movers. Any time a backscattering occurs, the event leaves its
footprint via a spin-flip in the nuclear memory. A reverse bias can now
be applied so that the current is opposite of the voltage bias in order
to extract work. We show below that the energy is supplied by the
thermal energy of the reservoirs. All this is reminiscent of a MD
operation wherein the MD predominantly backscatters the right movers
relative to the left movers, thus setting up a current between
reservoirs that are otherwise in equilibrium, while recording the
outcome in the nuclear spin memory (see Fig.~\ref{FIG:QIE_circuit}).
Under applied reverse bias, the MD/QIE harvests heat to convert it to
electrical work.

\subsection{Polarization dynamics and induced current}


We now quantify our model. The dynamics of the edge electrons are modified 
in the presence of nuclear spins via the spin-flip scattering which 
can be calculated using Fermi's golden rule in terms of the electron density matrix $\rho(r,r')$. 
For ease of notation, we
now focus on the bottom edge, where all right moving electrons have
spin up and left movers have spin down. The top edge results can be
obtained by substituting: spin down $\rightarrow$ spin up and vice
versa. Then, the scattering rate from right to left (with accompanying
nuclear spin flips) between $x$ and $x+\Delta x$ is given
by~\cite{REF:Lunde2012, REF:DelMaestro2013}:
\begin{eqnarray}\label{EQN:SpinFlipRate}
\Gamma_{-+}(\epsilon,x)= \frac{\gamma_0}{\hbar}\, N_\downarrow(x) \, f_+(\epsilon,x)\big(1-f_-(\epsilon,x)\big),
\end{eqnarray}
where $N_\downarrow(x)$ is the number of nuclear down spins between $x$
and $x+\Delta x$, $\gamma_0 \equiv \lambda^2/8\pi\hbar^2 v_F^2$ is a
dimensionless effective electron spin-nuclear spin interaction
strength. Here, $f_{\pm}(\epsilon,x)$ are the distributions of
right[left] movers with energy $\epsilon$ at position $x$ along the
edge 
and are given in terms of the Wigner transform of the density matrix $\rho_\pm(r,r')$ of the right and left moving electrons
as: 
\begin{equation}
f_\pm(\epsilon,x)= \int dr \rho_\pm(x+r/2,x-r/2) {\rm e}^{\pm i\epsilon r /\hbar v_F}.
\end{equation}

\subsubsection{Nuclear polarization dynamics}

The effect of spin-flip scattering on the nuclear spins is given by the
rate equation:
\begin{equation}\label{EQN:NucSpinDynamics}
\frac{dN_\uparrow(x)}{dt} = \int d\epsilon \big(\Gamma_{-+}(\epsilon, x)-\Gamma_{+-}(\epsilon, x)\big).
\end{equation}
We find it useful to define the mean polarization $m(x) \equiv
\frac{N_\uparrow(x) - N_{\downarrow}(x)}{2 (N_\uparrow(x)
+N_{\downarrow}(x))}$, whose time rate of change is
\begin{equation}\label{EQN:PolarizationDynamics}
\frac{dm(x)}{dt}=\gamma_0 \Gamma_B (x) - m(x) \gamma_0 \Gamma_T (x)
\end{equation}
with $\hbar \, \Gamma_B(x) = \int d\epsilon \, \big(f_+(\epsilon,
x)-f_-(\epsilon, x)\big)/2$ and $\hbar \, \Gamma_T(x) = \int d\epsilon
\, \big(f_+(\epsilon, x)+f_-(\epsilon, x)-2f_+(\epsilon,
x)f_-(\epsilon, x)\big)$. For the rest of the paper we focus on a short
edge, where the weak dependence of $f_\pm$ and $m$ on $x$ can be
ignored. Hence, we approximate $m(x)$ by its leading, $x$-independent,
order. We also approximate the distributions $f_+(\epsilon, x)$ of the
right movers and $f_-(\epsilon,x)$ of the left movers by the Fermi
distributions $f^0_{L}(\epsilon)$ and $f^0_{R}(\epsilon)$ of the
reservoirs from which they originate~(see Appendix \ref{SECT:Appendix_A}). We
therefore get
\begin{align}\label{EQN:GammaBGammaT}
\hbar\Gamma_B &= (\mu_L-\mu_R)/2,\nonumber\\
\hbar\Gamma_T &= (\mu_L-\mu_R)\coth\big(\frac{\mu_L-\mu_R}{2 k_B T}\big),
\end{align}
where $\mu_{L}[\mu_{R}]$ is the chemical potential of the left[right]
reservoir. We now use these expressions in  Eq.
\eqref{EQN:PolarizationDynamics} to obtain the time dependence of the
polarization:
\begin{align}\label{EQN:Polarization}
m(t)&=(m_0-\bar{m}){\rm e}^{-t/\tau_m}+\bar{m},
\end{align}
where $m_0$ is the initial mean polarization and $\bar{m} \equiv
\Gamma_B/\Gamma_T=(1/2) \tanh\big(\frac{\mu_L-\mu_R}{2 k_B T}\big)$ is
defined to be the target mean polarization and $\tau_m = 1/\gamma_0
\Gamma_T$ is the characteristic time scale for nuclear polarization
dynamics.

\subsubsection{Electron dynamics and induced current}

We now calculate the total current. The distribution functions obey the
Boltzmann-like equation for the bottom edge:
\begin{equation}
\partial_tf_\pm= \pm\big(\Gamma_{+-}(\epsilon, x)-\Gamma_{-+}(\epsilon, x)\big)\,\nu(0)^{-1}\mp v_F\partial_xf_{\pm},
\end{equation}
where $\nu(0)=L/2 \pi\hbar v_F$ is the density of states of the edge
electrons. We assume that the nuclear polarization $m$ is changing
slowly and seek a steady state solution. Then the distributions obey:
\begin{align}\label{EQN:Del_x_of_f_pm}
\partial_xf_\pm &=\big(\Gamma_{+-}(\epsilon, x)-\Gamma_{-+}(\epsilon, x)\big)\,\big(v_F\nu(0)\big)^{-1}\nonumber\\
				&\equiv\Gamma[f_+,f_-]
\end{align}
For short edges ($\Gamma[f_+,f_-]\, L \ll 1$), we expand in gradients
of the distribution functions. At the leading order, we obtain a linear
position dependence:
\begin{equation}\label{EQN:ExpandedDistFunc}
f_\pm=f^0_{L(R)}(\epsilon)+\Gamma[f^0_{L}(\epsilon),f^0_{R}(\epsilon)]\,(x\pm L/2).
\end{equation}
We then obtain the total current (see Appendix \ref{SECT:Appendix_A}):
\begin{equation}\label{EQN:Itot}
I_{tot}= \frac{e}{h} \int d\epsilon (f_+-f_-)= \frac{e^2}{h}V -e N\gamma_0 ( \Gamma_B -
m \Gamma_T).
\end{equation}
We identify and focus on two sources of current in the system in the
short edge regime: (i) the usual current $\frac{e^2}{h} V$ due to
voltage bias without the nuclear spin-flip interaction, and (ii) the
MD-induced current $-e N\frac{dm}{dt} = -e N\gamma_0 \, (\Gamma_B -
m\Gamma_T)$ due to the presence of nuclear polarization $m$. In the
latter case, a net backscattering current, caused by right-moving
up-spin electrons scattering to left-moving down-spin electron states,
is driven by a net negative nuclear spin and vice versa. We note that
the net polarization of the nuclear spins acts as a Maxwell's Demon:
The total current is nonzero for vanishing bias voltage, demonstrating
the ``Demon action'' that induces a current between two reservoirs at
equal temperature and chemical potential, while using the nuclear spins
as a memory resource.

\begin{figure}[tb]
\centerline{\includegraphics[width=0.95\linewidth]{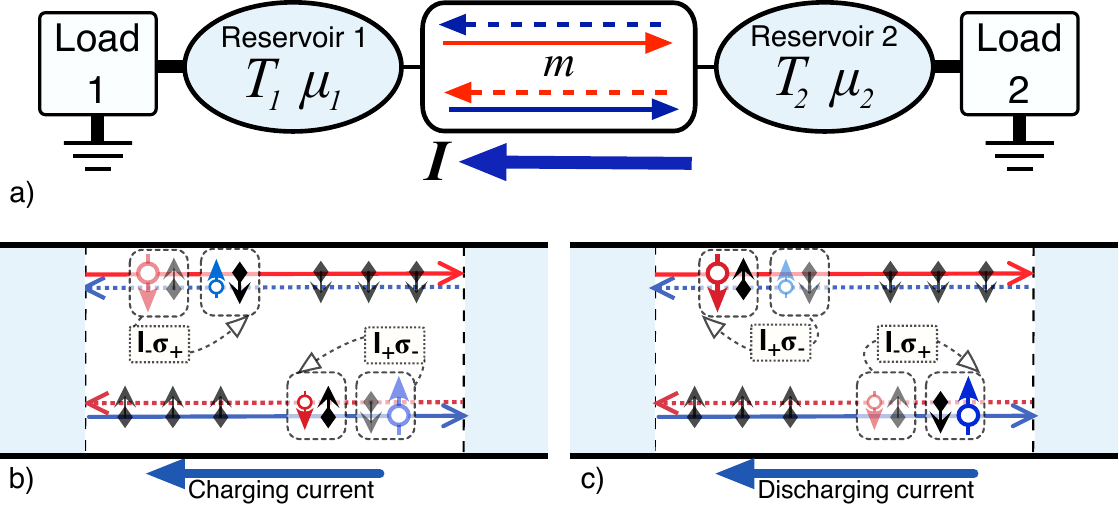}}
\caption{(Color online) a) QSHI based quantum information engine,
providing power to loads 1 and 2. Schematic description of b) the
charging and c) the discharging phase of the QIE. In the charging
phase, an applied bias current increases the number of right-movers
(solid lines) at the edges relative to left-movers (dashed lines). This
excess in turn creates a net nuclear spin polarization with opposite
values in each edge. In the discharging phase, even without external
bias, the net polarization of the nuclear spins increases the number of
right-movers, driving a net discharging current to the
left.}\label{FIG:QIE_circuit}
\end{figure}

\subsection{Generated power and extracted work}

\begin{figure}[tb]
\includegraphics[width=\linewidth]{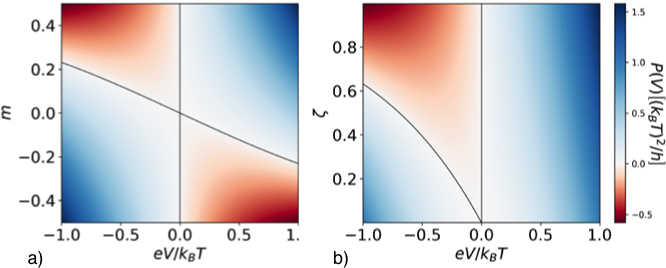}
\caption{(Color online)  Power, calculated using
Eq.~\eqref{EQN:ShortEdge_Power}, as a function of the applied voltage
and a) as a function of mean polarization $m$ with $\zeta = 1.0$ and b)
as a function of $\zeta$ with full polarization m = 0.5. The lines
divide the charging ($P>0$) and discharging ($P<0$) phases. Power can
have negative values for $V<0$ $(V>0)$ for a given mean polarization
$m>0$ $(m<0)$ (here, $e>0$), as an indication of the work extraction
phase. }\label{FIG:shortedge_power}
\end{figure}

In order to use the quantum information engine, we attach it to an
electrical circuit as in Fig.~\ref{FIG:QIE_circuit}. In this setup, the
QIE provides power to loads 1 and 2, which can be modeled by a
(reverse) bias voltage $V$. The power generated
(Fig.~\ref{FIG:shortedge_power}) is given by:
\begin{equation}\label{EQN:ShortEdge_Power}
P =\frac{eV}{h} \bigg(eV (1-\frac{\zeta}{2}) + \zeta\, \hbar\Gamma_T \,m \, \bigg),
\end{equation}
with $\zeta = 2\pi N\gamma_0$. For $eV<\frac{2\zeta \hbar\Gamma_T
m}{(\zeta-2)}$, we obtain $P<0$, indicating that the circuit is powered
by the QIE. (For $eV>\frac{2\zeta  \hbar\Gamma_T m}{(\zeta-2)}$, the
circuit is providing power to charge the nuclear spin resource).
We find the maximal work done by the nuclear spin resource in the weak
coupling/short edge limit by maximizing the power and integrating up to
the time when the power changes sign~\cite{FTN:Lifetime}:
\begin{equation}\label{EQN:ShortEdge_EnergyDensity}
W_{tot} = \alpha k_B T N^2 \gamma_0,
\end{equation}
where $\alpha$ is a parameter of $\mathcal{O}(1)$ (a detailed
calculation produces $\alpha = \pi/4$ for a work extraction under
constant voltage bias (see Appendix \ref{SECT:Appendix_B}). In this limit, the amount
of extracted work follows a quadratic scaling law that implies denser
storage than the conventional/expected linear scaling. Here, $T$ is the
operating temperature, limited by the bulk band gap of the QSHI. We
find that the inequality
\begin{equation}\label{EQN:2ndLawMain}
P +  k_B T \dot{S}_{\text{nuc}} \geq 0,
\end{equation}
where $S_{\text{nuc}}$ is the information entropy of the nuclear spin
subsystem , is satisfied in agreement with the second law of
thermodynamics (see Appendix \ref{SECT:Appendix_C}).

\subsection{Physical implementation}

We now discuss experimental feasibility of our MD implementation.
Systems featuring spin-momentum locked topological edge states have
been available to experiments for about a decade~\cite{REF:Hasan2010,
REF:Qi2011}.  Among these materials, systems with high nuclear spin
density generally provide high energy density. In addition, systems
with higher bulk bandgaps could be operated at higher temperatures,
again leading to higher energy densities (see
Eq.~\eqref{EQN:ShortEdge_EnergyDensity}). Systems that feature high
hyperfine interaction strength or low Fermi velocity provide high power
density and fast operation, thus can be utilized as
spin-supercapacitors. Assuming $N\sim 10^7$ and $\gamma_0 \sim
10^{-8}$, we estimate the equivalent energy density and power density
that can be stored in the device in the short edge limit to be $\sim 10
\textrm{kJ}/\textrm{kg}$ and $\sim 10 \textrm{MW}/\textrm{kg}$ (not
including overhead). On the other hand, systems with low interaction
strength (see Eq.~\eqref{EQN:SpinFlipRate}) due to high Fermi velocity
and/or suppressed hyperfine interaction can be utilized as spin
batteries that keep their polarization for long times. For example,
thin film flakes of 3D topological QSHI
$\textrm{Bi}_2\textrm{Te}_2\textrm{Se}$ (BTS221) feature a relatively
large Fermi velocity ($v_\textrm{F}\sim 10^6 m/s$)~\cite{REF:Xu2010},
which is two orders of magnitude larger than that of, say, InAs/GaSB
QWs ($v_\textrm{F}\sim 10^4 m/s$)~\cite{REF:Du2015}. Thus, BTS221
features a much smaller electron-nuclear spin-flip interaction strength
(therefore requiring large currents to write the spin memory) and
orders of magnitude longer memory retention times. In fact, recent
experimental work that uses thin film flakes of BTS221 observed days
long polarization retention times~\cite{REF:Tian2017}.

We next consider InAs/GaSb QW structures as an example. These QWs have
a smaller Fermi velocity $v_F$~\cite{REF:Konig2008, REF:Du2015} and
higher nuclear spin density compared to, for example, HgCdTe
QWs~\cite{REF:Lunde2013}. This hints to a larger $N \gamma_0$ in
InAs/GaSb QWs and therefore to a faster operation and higher energy
density. We note that in these QWs, the electrons have spin $\pm 1/2$
but the holes have spin $\pm 3/2$ whose coupling to the spin-flip
interaction requires a higher order process~\cite{REF:Lunde2013}. The
nuclear spin density in these QWs, as well as the effective electron
spin--nuclear spin coupling strength, could possibly be further
adjusted by magnetic impurity doping, providing a design freedom that
might prove useful for different functionalities of the QIE.

\section{Conclusion}


In summary, we have described a Maxwell's Demon system that utilizes
the spin-flip interaction between helical edge states and nuclear spins
in quantum spin Hall topological insulators. Available nuclear or
magnetic impurity spins can be utilized as a Maxwell's Demon memory to
harvest work from thermal energy of the reservoirs. We also showed how
to erase the memory and thus ``charge'' the system by applying a
voltage bias. Erasing the memory (or polarizing the spin subsystem)
requires dissipation of heat by an amount at least $k_B T \ln 2$ per
bit, in agreement with the Landauer's principle and the second law.
Estimates of equivalent work that can be extracted show that
power/energy densities that exceed existing supercapacitors are
achievable.

\begin{acknowledgments}


We thank \.{I}. \.{I}.
Kaya , J. Vaccaro and \"{O}. M\"{u}stecapl{\i}o\u{g}lu for discussions. We especially thank N. Allen for inspiration as
well as many fruitful discussions throughout this work. IA is a member
of the Science Academy--Bilim Akademisi--Turkey; BP and AMB thank The
Science Academy--Bilim Akademisi--Turkey for the use of their
facilities. This research was supported by a Lockheed Martin
Corporation Research Grant. IA and BP also acknowledge support by COST
Action MP1209.

\end{acknowledgments}

\appendix

\section{Mean Polarization Dynamics and Electric Current}\label{SECT:Appendix_A}


In this appendix, we focus on the time dependence of the nuclear mean
polarization and later obtain electric current in the short edge limit.
We first define the mean polarization at position $x$ per edge,
\[ m(x) \equiv \frac{N_\uparrow(x)
-N_{\downarrow}(x)}{2 N(x)}
,
\]
where $N(x) =  N_{\uparrow}(x) + N_{\downarrow}(x)$ is the number of
nuclei at position $x$ and $x + \Delta x$ per edge participating in the
spin-flip interactions. With this definition fully polarized nuclear
spins have $m=\pm 1/2$.

The dynamics of the mean polarization, given in
Eq.~\eqref{EQN:PolarizationDynamics} and repeated below, is obtained
from Eq.~\eqref{EQN:SpinFlipRate} and \eqref{EQN:NucSpinDynamics}:
\begin{equation}
\frac{dm(x)}{dt}=\gamma_0 \Gamma_B (x) - m(x) \gamma_0 \Gamma_T (x) \tag{\ref{EQN:PolarizationDynamics}}.
\end{equation}
As mentioned in the main text, $\Gamma_B(x)$ and $\Gamma_T(x)$ is given
as:
\begin{align}\label{EQN:Splmtl:GammaBGammaT_PosDep}
\Gamma_B(x) &= \int \frac{d\epsilon}{\hbar} \frac{f_{+} - f_{-}}{2}\nonumber\\
\Gamma_T(x) &= \int \frac{d\epsilon}{\hbar} \big( f_{+} + f_{-} - 2f_{+}f_{-}\big),
\end{align}
where we suppressed the energy and position dependence of the
distribution functions $f_\pm(\epsilon, x)$. Note that current
conservation requires $\Gamma_B(x)$ to be $x$-independent.

For short edges we have $\Gamma[f_+,f_-]\,L \ll 1$, and
$N_{\uparrow(\downarrow)}(x)$ and $m(x)$ have only a weak dependence on
$x$. Performing a gradient expansion, we first approximate
$N_{\uparrow(\downarrow)}(x)$ and $m(x)$ with their leading,
$x$-independent, terms. We next approximate the distributions $f_\pm$
with the Fermi distributions of the reservoirs $f^0_{L(R)}$ from which
they originate. We now evaluate the integrals in
Eq.~\eqref{EQN:Splmtl:GammaBGammaT_PosDep} and obtain
\begin{align}
\Gamma_B &= (\mu_L-\mu_R)/2\hbar,\nonumber\\
\Gamma_T	&= \frac{(\mu_L - \mu_R)}{\hbar} \coth{\left(\frac{\mu_L - \mu_R}{2kT}\right)},
\end{align}
in agreement with Eq.~\eqref{EQN:GammaBGammaT}. We note that in this
approximation, $\Gamma_B$ is proportional to the applied bias
$\mu_L-\mu_R$, hence it vanishes for zero applied voltage. We also note
that $\Gamma_T\geq 0$.

We now focus on the total current. In the short edge limit, the
distribution functions of the right and left movers within the edge in
question are given by Eq.~\eqref{EQN:ExpandedDistFunc}. We then obtain
the total current as
\begin{align*}
I_{tot}&= \frac{e}{h} \int d\epsilon \big(f_+-f_-\big)\nonumber\\
&=  \frac{e}{h} \int d\epsilon \bigg[\big(f_L^0 - f_R^0\big) - h\big(\Gamma_{-+}(\epsilon)-\Gamma_{+-}(\epsilon)\big) \bigg] \nonumber\\
&=\frac{e^2}{h}\,V - eN \frac{d m}{dt},
\end{align*}
consistent with Eq.~\eqref{EQN:Itot}. Here, $f_+$ [$f_-$] are the (in
general $x$-dependent) distribution functions of the right [left]
movers in the given edge, and $f_L^0$ [$f_R^0$] are the (Fermi)
distributions of the lead from which the right [left] movers originate.
In the last line, we used Eq. \eqref{EQN:NucSpinDynamics} in the short
edge limit. We identify the first term on the right hand side as the
current due to the usual voltage bias ($I_\text{bias}$) and the second
term as the induced current due to Maxwell's demon effect ($I_{MD}$).

In order to see the mean polarization dependence of the total current
in the short edge limit, we substitute the explicit forms of $\Gamma_B$
and $\Gamma_T$ given in Eq.~\eqref{EQN:GammaBGammaT} into
Eq.~\eqref{EQN:Itot} to obtain:
\begin{align}
I_{\text{tot}}\left(t\right)	= \frac{e^2}{h}\,V\, \bigg[\left(1-\frac{\zeta}{2} \right)
								+ m(t)\,\zeta \, \coth{\left(\frac{eV}{2kT}\right)}\bigg],
\end{align}
where we  defined a dimensionless quantity $\zeta = 2\pi N\gamma_0$,
which is a rough measure of the interaction strength over the whole
wire per edge. We see that in the limit of vanishing voltage bias,
total current is not zero if $m(t)$ is nonzero. This behaviour persists
even if the temperature or the chemical potential of both of the
reservoirs are equal, demonstrating the pure entropy-driven current.


\section{Work Extraction and Heat Dissipation}\label{SECT:Appendix_B}


We calculate the power absorbed/generated by QIE under fixed applied
voltage bias as follows:
\begin{align}\label{EQN:Splmtl:Power}
P(t)	&= I_{\text{tot}}(t)\,V\nonumber\\
		&= \frac{eV}{h}\Big[eV(1-\frac{\zeta}{2}) + \zeta\, m(t)\, \hbar \Gamma_{T}\Big]\nonumber\\
		&= \frac{eV}{h}\Big[eV + \zeta  (m_0-\bar{m}){\rm e}^{-t/\tau_m}\hbar \Gamma_{T}\Big],
\end{align}
where in the last line we used Eq.~\eqref{EQN:Polarization} and $\bar{m}\hbar
\Gamma_{T} = \frac{eV}{2}$.

{\it Charging cycle.} We would like to find the amount of heat dissipated while we charge the
device. Starting from totally unpolarized nuclear spins ($m_0 = 0$) and
using Eq. \eqref{EQN:Polarization} and Eq.~\eqref{EQN:Splmtl:Power}, we
get:
\begin{align}\label{EQN:Splmtl:ChargingPower}
P(t)&=\frac{eV_C}{h}\Big[eV_C - \zeta \bar{m}{\rm e}^{-t/\tau_m}\hbar \Gamma_{T}\Big],\nonumber\\
&=\frac{eV_C}{h}\Big[eV_C - eV_C \frac{\zeta }{2}{\rm e}^{-t/\tau_m}\Big].
\end{align}
As shown in Eq. \eqref{EQN:Polarization}, the amount of time to reach
the target mean polarization is infinitely long. Instead, we charge the
device up to a fraction of full polarization $m = \frac{\kappa}{2}$
where $\kappa$ is a value we later choose depending on the application
and whether we intend to maximize power or efficiency. Using
$\frac{\kappa}{2} = \bar{m}(1-e^{-\bar{t}/\tau_m})$, we obtain the
following for the amount of time $\bar{t}$ to reach the specified
target mean polarization:
\begin{align}
\bar{t} = -\tau_m \ln{\left(1-\frac{\kappa}{2\bar{m}}\right)}.
\end{align}
We then get the dissipated heat by integrating the power up to
$\bar{t}$:
\begin{align}
W_{C}(V_C)	&= \int_0^{\bar{t}} \frac{eV_C}{h}\Big[eV_C  - \frac{\zeta}{2} {\rm e}^{-t/\tau_m}eV\Big]  \nonumber\\
			&= \frac{e^2V_C^2}{h}\tau_m \Big[-\ln\left({1-\frac{\kappa}{2\bar{m}}}\right) - \frac{\zeta}{2}\frac{\kappa}{2\bar{m}}\Big]\nonumber\\
			&=\frac{eV_C}{2\pi\gamma_0}\tanh\left( \frac{eV_C}{2k_B T}\right) \nonumber\\
			&\qquad \times \, \Big[\ln\left(\frac{2\bar{m}}{2\bar{m}-\kappa}\right) - \frac{\zeta}{2}\frac{\kappa}{2\bar{m}}\Big].
\end{align}
Note that $0\leq 1- \kappa/2\bar{m}<1$. This condition gives us an
lower bound on the applied voltage:
\begin{align}
\label{EQN:Splmtl:chargingvoltage}
V_C &\geq \frac{k_B T}{e} \ln\left(\frac{1 + \kappa}{1-\kappa}\right).
\end{align}


{\it Discharging cycle.} For the next step in the engine cycle, we
apply a reverse (discharging) bias, $V_D<0$, and we would like to find
the time $t^{*}$ at which $P(t)$ changes sign. Using
Eq.~\eqref{EQN:Splmtl:Power}, we obtain:
\begin{align}
t^{*} &= \tau_m \ln\left[\zeta\left( \frac{1}{2} + m_0 \coth\left( \frac{|eV_D|}{2k_B T}\right) \right)\right]
\end{align}
We then integrate the power up to $t=t^*$ to obtain the work done at
fixed voltage:
\begin{align}
W_{D}(V_D)	&= \int_0^{t^{*}} \frac{|eV_D|}{h}\Big[|eV_D| - \zeta  (m_0-\bar{m}){\rm e}^{-t/\tau_m}\hbar \Gamma_{T}\Big]\nonumber\\
			&= \frac{e^2 V_D^2}{h}t^{*} \nonumber\\
			&\qquad + \frac{|eV_D|}{h}\tau_m\zeta(m_0-\bar{m})\hbar\Gamma_T ({\rm e}^{-t^{*}/\tau_m}-1).
\end{align}
Inserting $t^{*}$ into the equation above and and using the relation
$\bar{m}\hbar \Gamma_{T} = -\frac{|eV_D|}{2}$, we get:
\begin{align}
W_{D}(V_D) &= \frac{e^2 V_D^2}{h}\tau_m \bigg[\ln\left(\frac{\zeta}{2} + m_0\,\zeta \, \coth\left( \frac{|eV_D|}{2k_B T}\right) \right)\nonumber\\
&+ 1 - \left(\frac{\zeta}{2} + m_0 \,\zeta\,\coth\left( \frac{|eV_D|}{2k_B T}\right)\right)\bigg].
\end{align}
We finally take  the polarization reached at the end of the charging
cycle, $m_0 = \frac{\kappa}{2}$, as the initial polarization for the
discharging cycle to finally obtain
\begin{align}
\label{EQN:Splmtl:WorkExt}
W_{D}(V_D) 	&= \frac{|eV_D|}{2\pi \gamma_0} \tanh\left( \frac{|eV_D|}{2k_B T}\right) \nonumber\\
			&\qquad \times \, \Bigg[\ln\left( \frac{\zeta}{2} +  \frac{\kappa \, \zeta}{2} \,\coth\left( \frac{|eV_D|}{2k_B T}\right) \right)\nonumber\\
			&\qquad + 1 -\left( \frac{\zeta}{2} +  \frac{\kappa \, \zeta}{2} \,\coth\left( \frac{|eV_D|}{2k_B T}\right) \right)\Bigg].
\end{align}
In order to extract work from the nuclear spin polarization, one has to
make sure that $t^* > 0$, which gives us an upper bound on the applied
voltage:
\begin{align}
\label{EQN:Splmtl:DischargeBound}
|V_D| \leq  \frac{k_B T}{e} \ln\left(\frac{2 - \zeta(1- \kappa)}{2- \zeta(1+ \kappa)}\right).
\end{align}


{\it Maximum Work Extraction}: Eq.~\eqref{EQN:Splmtl:DischargeBound}
suggests that, in the short edge limit, work extraction is only
possible when the applied reverse bias is smaller than the thermal
energy. We therefore consider the $2k_B T \gg |eV_D|$ case and
approximate Eq.~\eqref{EQN:Splmtl:WorkExt} as follows:
\begin{align}\label{EQN:Splmtl:WorkApp}
W_{D}(V_D)	&\approx \frac{|eV_D|^2}{4k_B T\pi \gamma_0} \Bigg[\ln\Bigg(\frac{\zeta}{2}\Big( 1 +  \kappa\frac{2k_B T}{|eV_D|} \Big)\Bigg)\nonumber\\
			& + 1 -\frac{\zeta}{2}\Bigg( 1 +  \kappa \frac{2k_B T}{|eV_D|} \Bigg)\Bigg].
\end{align}
The maximum work that can be extracted from QIE can be obtained by
maximizing Eq.~\eqref{EQN:Splmtl:WorkExt} with respect to applied
reverse bias $V_D$. We neglect the logarithmic term in
Eq.~\eqref{EQN:Splmtl:WorkExt} and we find the applied reverse bias
that maximizes the amount of extracted work:
\begin{align}\label{EQN:Splmtl:MaxVoltage}
|eV_D^*| =\frac{k_B T\zeta \kappa}{\left(2-\zeta\right)}.
\end{align}
Plugging $V_D^*$ into Eq.~\eqref{EQN:Splmtl:WorkApp} and choosing
maximum initial polarization $\kappa = 1$, we get the maximum work that
can be extracted under constant voltage bias in the short edge limit
as:
\begin{align}\label{EQN:Splmtl:WorkExtMax}
W_{tot} \simeq \frac{\pi}{4} k_B T N^2 \gamma_0.
\end{align}


\section{Information Entropy and the Second Law of Thermodynamics}\label{SECT:Appendix_C}


%
The second law of thermodynamics in our context can be restated as
\begin{equation}\label{EQN:2ndLaw1}
\beta \,P +  \frac{d}{dt}{S}_{\text{nuc}} \geq 0,
\end{equation}
where $\beta\equiv (k_B T)^{-1}$ and $P$ is the power dissipated at
(extracted from) the reservoirs, $T$ is the ambient temperature and
$S_{\text{nuc}}$ is the information entropy of the nuclear spin
subsystem. As we extract work using nuclear spins as a memory resource
$(P< 0)$, we see that the information entropy of the nuclear spin
subsystem has to increase. In the reverse process in which we erase the
memory $(P>0)$, the information entropy of the nuclear spin subsystem
decreases, which corresponds to the Landauer's erasure principle.
\begin{figure}
\includegraphics[width=.99\columnwidth]{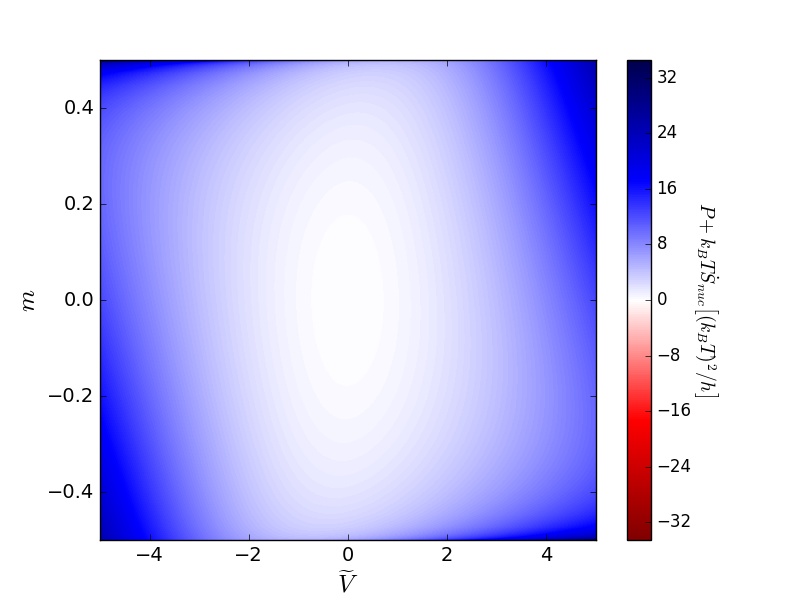}
\caption{(Color online) $P+k_B T\dot{S}_\text{nuc}$ as a function of
$\widetilde{V}$ and $m$ for $\zeta=1$.  Note that $P+k_B
T\dot{S}_\text{nuc}\geq 0$, in agreement with the second law.}
\label{FIG:2ndLaw}
\end{figure}

For our system, i.e.~QIE in the short edge limit, we use Eq.
\eqref{EQN:Splmtl:Power} and obtain
\begin{align}\label{EQN:2ndLaw2}
\beta \, P + \dot{S}_{\text{nuc}} &= \frac{1}{\beta h}\,\big[ \widetilde{V}^2+ \zeta \widetilde{V}(\widetilde{V}+X) \nonumber\\
&\qquad \times (m\coth{\frac{\widetilde{V}}{2}} - \frac{1}{2}) \big]\geq 0,
\end{align}
where $\widetilde{V} \equiv \beta \,eV$ and $X = \ln\big(
\frac{1+2m}{1-2m} \big)$, in agreement with the second law of
thermodynamics (see Fig.~\ref{FIG:2ndLaw}).


\end{document}